\documentclass[reprint,preprintnumbers,amsmath,amssymb,bibnotes]{revtex4-2}

\usepackage{graphicx}% Include figure files
\usepackage{dcolumn}% Align table columns on decimal point
\usepackage{bm}% bold math
\usepackage{cancel} % cross out terms
\usepackage{color}

\usepackage{url}
\usepackage{lineno,hyperref}

\usepackage{float}
\restylefloat{figure}
\usepackage{epstopdf}

\usepackage{CJK}

\newcommand{\req}[1]{(\ref{#1})}

\def\fc#1#2{\frac{#1}{#2}}
\def\h{\frac{1}{2}}
\newcommand{\nwc}{\newcommand}
\nwc{\ba}  {\begin{array}}
\nwc{\ea}  {\end{array}}
\nwc{\bdm} {\begin{displaymath}}
\nwc{\edm} {\end{displaymath}}

\nwc{\bea} {\begin{equation}\ba{lcl}}
\nwc{\eea} {\ea\end{equation}}

\nwc{\be} {\begin{equation}}
\nwc{\ee} {\end{equation}}

\nwc{\bda} {\bdm\ba{lcl}}
\nwc{\eda} {\ea\edm}

\nwc{\bc}  {\begin{center}}
\nwc{\ec}  {\end{center}}

\nwc{\ds}  {\displaystyle}

\nwc{\nn} {\nonumber}
\nwc{\nnn} {\nonumber \vspace{.2cm} \\ }
\nwc{\ra}{\rightarrow}
\nwc{\lra}{\longrightarrow}
\def\lf{\left}\def\ri{\right}

\nwc{\p} {\partial}
\def\Fc{{\cal F}}
\def\ap{\alpha'}

\def\Mc{{\cal M}}

\def\ov{\overline}

\def\IC{{\bf C}}

\def\IZ{{\bf Z}}

\def\Fc{{\cal F}}
\def\Sc{{\cal S}}

\def\si{\sigma}

\begin{document}
\begin{CJK*}{GB}{} % Use default fonts from CJK (see below)
\preprint{MPP--2023--241}
\title{One--loop Double Copy Relation in String Theory
%  \\ (One--Loop KLT Relation)
}
\author{S. Stieberger}
\affiliation{Max--Planck--Institut f\"ur Physik, Werner--Heisenberg--Institut, 80805 M\"unchen, Germany}
\begin{abstract}
We discuss relations between closed and open string  amplitudes at one--loop.
While at tree--level these relations are known as Kawai--Lewellen--Tye (KLT) and/or double copy relations, here we investigate how such relations are manifested at one--loop. Although, we find  examples of one--loop closed string amplitudes  that can strikingly  be written as sum over squares of one--loop open string amplitudes, generically the one--loop closed string amplitudes
assume a form reminiscent  from the one--loop doubly copy structure of gravitational amplitudes involving a loop momentum. 
This double copy structure  represents the one--loop generalization of the KLT relations. 

\end{abstract}
\maketitle
\end{CJK*}

\section{Introduction}

The famous Kawai--Lewellen--Tye (KLT) relations  express a tree--level closed string amplitude as a weighted sum over squares of tree--level open string amplitudes~\cite{Kawai:1985xq}.
Since the lowest mode of the closed superstring is a graviton and that of the open superstring a gluon the aforementioned relation gives rise to a  gauge/gravity correspondence  linking gravity and gauge amplitudes at the perturbative tree--level.
This connection  has far reaching consequences after elevating it to the double copy (DC)  
conjecture \cite{Bern:2008qj}. At an abstract level the KLT relations provide
 a way of computing tree--level closed string world--sheet integrals by reducing them to open string integrals. 
At the technical level the latter statement means that  a complex sphere integral can be expressed in terms of a product of two iterated real integrals.
While conjectures based on generalized unitarity for perturbative quantum gravity as a DC structure  exist for field theory loop--level \cite{Bern:2010ue}, only recently a one--loop  analog has been found in string theory \cite{Stieberger:2022lss}. In 
\cite{Stieberger:2022lss} a one--loop extension of the KLT relations has been derived and in this work we elaborate  on the underlying DC structure.

%%  involves theta--functions on torus.

\section{Closed vs. open string amplitudes}

Closed string amplitudes are described by  integrals over compact Riemann surfaces without boundaries and open string amplitudes are formulated on world--sheets with boundaries. 
Surfaces with boundaries are obtained from manifolds without boundaries  by involution.
While closed string vertex positions are integrated over the full manifold those of open strings are integrated along boundaries only.
To find relations between closed and open string amplitudes an analytic continuation of each complex closed string coordinate is performed to  split the latter into a pair of two real coordinates. The  latter describe open string vertex positions located at the boundaries of the underlying world--sheet.
At the mathematical level relations between closed and open string amplitudes are subject
to holomorphic properties of the string world--sheet and underlying monodromy relations, cf. \cite{Stieberger:2009hq,Bjerrum-Bohr:2009ulz} for tree--level and \cite{Hohenegger:2017kqy,Tourkine:2016bak,Casali:2019ihm} for one--loop.
In fact, while these relations are  formulated on surfaces with boundaries, 
they can be extended to surfaces without boundaries \cite{Stieberger:2022lss}.

%%\vskip2cm
\ \\
\centerline{\bf A. Complex sphere integral}\\

Closed string tree--level $n$--point amplitudes are described by an integral over the moduli space of 
$n$ marked points on the sphere $\IC$.
For $n=4$ we have the  integral
 \begin{align}
M^{closed}_{4;0}&:=\int_\IC
 d^2z\ |z|^{2\ap s-2}\ |1-z|^{2\ap u}\nonumber\\
 &=\fc{\Gamma(\ap s)\Gamma(\ap t)\Gamma(\ap u)}{\Gamma(1-\ap s)\Gamma(1-\ap t)\Gamma(1-\ap u)},
\end{align}
with $s\!+\!t\!+\!u\!=\!0$ and referring to a four--point closed string tree--level amplitude to be specified below.
%%%with the bosonic closed string tree--level Green function
%%%\be\label{BGtree}
%%%G^{(0)}_B(z)=\ln|z|^2
%%%\ee
On the other hand, with the corresponding open string disk integrals
\begin{align}
A^{open}_{4;0}&:=\int_0^1 d\xi\ \xi^{\ap s-1}\ (1-\xi)^{\ap u}=\fc{\Gamma(\ap s)\Gamma(\ap u+1)}{\Gamma(1-\ap t)},\label{Open1}\\
\tilde A^{open}_{4;0}&:=\int_1^\infty d\eta\ \eta^{\ap t-1}\ (\eta-1)^{\ap u}=\fc{\Gamma(\ap t)\Gamma(\ap u+1)}{\Gamma(1-\ap s)},
\end{align}
we have:
 \be\label{KLT4}
M^{closed}_{4;0}= \sin(\pi \ap u)\ A^{open}_{4;0}\ \tilde A^{open}_{4;0}\ .
 \ee

Actually, \req{Open1} enters the open superstring subamplitude describing the scattering of four (massless) gluons 
\be
A^{open}_{4;0}(1,2,3,4)=\fc{t_8}{u}\ A_{4;0}^{open}
\ee
with canonical color ordering $(1,2,3,4)$. With the four external gluon momenta $p_i$ (subject to the massless condition  $p_i^2=0$)  the three parameters $s,t,u$ refer to the kinematic invariants $s\!=\!2\ap p_1p_2, t\!=\!2\ap p_1p_3, u\!=\!2\ap p_1p_4$, respectively. Likewise, the four graviton closed superstring amplitude is given by: 
\be\label{closed4}
{\cal M}^{closed}_{4;0}=\fc{t_8\tilde t_8}{u^2}\ M^{closed}_{4;0}.
\ee
Thus we have the gravity--gauge relations or four--point KLT relation:
\be\label{gravity4}
{\cal M}^{closed}_{4;0}=\sin(\pi \ap u)\; A^{open}_{4;0}(1,2,3,4)\; \tilde A^{open}_{4;0}(1,3,2,4).
\ee

\begin{widetext}
Similar results can be stated for higher $n$ or massive states:
\bea\label{GravKLT}
\Mc_{n;0}^{closed}&=&\ds\kappa^{n-2}\ \sum_{\sigma,\rho\in S_{n-3}}A^{open}_{n;0}(1,\sigma(2,3,\ldots,n-2),n-1,n)\\[5mm]
&&\ds\times\ \ \Sc[\rho|\sigma]_{p_1}\
  \tilde A^{open}_{n;0}(1,\rho(2,3,\ldots,n-2),n,n-1),
  \eea
  involving the KLT--kernel $\Sc[\rho|\sigma]_{p_0}$ (intersection matrix). Generically, the latter is defined as a symmetric $k!\times k!$--matrix with its rows and columns corresponding to the orderings $\si \equiv \{\si(2),\ldots,\rho(k)\}$ and
$\rho \equiv \{\rho(1),\ldots,\rho(k)\}$, respectively. 
For given (cyclic)  orderings $\rho,\sigma\in S_{k}$  and a reference momentum $p_0$ one defines  the KLT kernel as    \cite{Kawai:1985xq,Bern:1998sv,Bjerrum-Bohr:2010pnr}
\be\label{kernel}
\Sc[\sigma|\rho]_{p_0}:=\Sc[\, \si(1,\ldots,k) \, | \, \rho(1,\ldots,k) \, ]_{p_0} =\prod_{t=1}^{k} \sin\Bigg(\pi  \ap \!\Big[p_0 p_{t_\si}+\sum_{r<t}p_{r_\si}p_ {t_\si} \theta(r_\si,t_\si)\Big]\Bigg),
\ee
with $j_\si=\si(j)$ and  $\theta(r_\si,t_\si)=1$
if the ordering of the legs $r_\si,t_\si$ is the same in both orderings
$\si(1,\ldots,k)$ and $\rho(1,\ldots,k)$, and zero otherwise. For the case at hand  \req{GravKLT} we
have $p_0=p_1$ and $k=n-3$. Finally, $\kappa$ is the gravitational coupling  related to Newton's constant via $\kappa^2 = 32\pi^2G_N$.
\end{widetext}

%\subsection{Complex torus integral}
\ \\
\centerline{\bf B. Complex torus integral}\\

\begin{figure}[H]
\centering
 \includegraphics[width=0.35\textwidth]{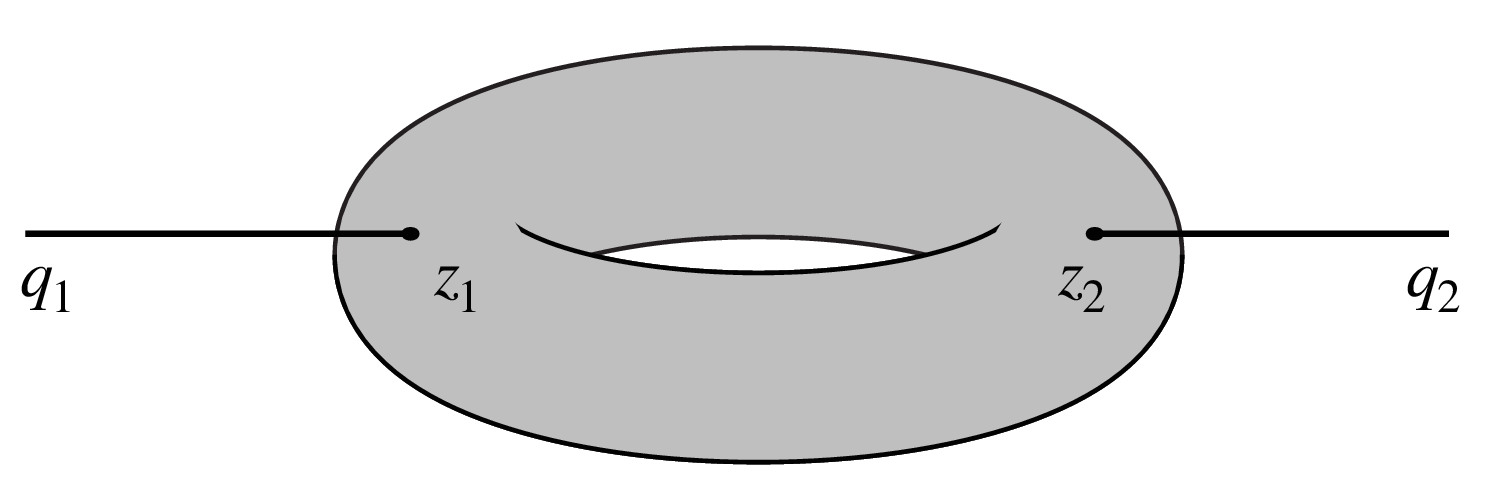}
\caption{One--loop amplitude with two massive closed strings $q_i^2\neq 0$.}
\label{Mass}
\end{figure}

Closed string one--loop $n$--point amplitudes are described by an integral over the moduli space of 
$n$ marked points on the elliptic curve ${\cal T}$. Let us discuss the one--loop torus integral $(n=2)$
\begin{align}
\widehat M^{closed}_{2;1}&:=\int_{\cal T}d^2z\; e^{2G^{(1)}(z,\tau)}\nonumber\\
&=2\tau_2^{\h}\lf|\fc{\theta_3\lf(2\tau\ri)}{\eta^6}\ri|^2+2\tau_2^{\h}\lf|\fc{\theta_2\lf(2\tau\ri)}{\eta^6}\ri|^2,\label{Marcus}
\end{align}
referring to a specific two--point closed string one--loop amplitude to be specified below. Above, we have introduced the bosonic one--loop Green function
\be\label{BG}
G^{(1)}(z,\tau)=\ln\lf|\fc{\theta_1(z,\tau)}{\theta_1'(0,\tau)}\ri|^2-2\pi\fc{(\Im z)^2}{\Im \tau}\ ,
\ee
the odd Riemann theta--function
\be\label{theta1}
\theta_1(z,\tau)=q^{\frac{1}{8}}\sum_{n\in\IZ}(-1)^n q^{\h n(n+1)}\sin[\pi(2n+1)z],
\ee
and $\eta=q^{1/24}\prod_{n\geq 1}(1-q^n),\; q=e^{2\pi i\tau}$. The complex torus coordinate is parameterized  as $z=x+\tau y$, with $x,y\in(0,1)$ and the measure
$d^2z=\tau_2dxdy$.
Actually, the integral \req{Marcus} describes a one--loop two--point amplitude. In superstring theories the latter and thus mass shifts vanish for massless 
states, while they do not vanish for massive states, cf. Fig. \ref{Mass}
\noindent
The two--point amplitude appears as residuum at the first massive level in the factorization of a four--point one--loop amplitude on its  double--pole in the $s$--channel accounting for the mass renormalization in superstring theory  \cite{Marcus:1988vs}.
The integral \req{Marcus} computes the mass correction $\delta m^2=\Mc^{closed}_{2;1}$ of the least massive string state in type II superstring theory \cite{Marcus:1988vs}
\begin{align}
\Mc^{closed}_{2;1}&=\delta^{(d)}(q_1+q_2) \int\fc{d^2\tau}{\tau_2}\; \tau_2^{-4}\nonumber\\
&\times\int_{\cal T}\!d^2z\; e^{-\fc{\ap}{2}q_1^2G^{(1)}(z,\tau)}\;,
\end{align}
subject to momentum conservation $q_1+q_2=0$ and the on--shell condition for the first massive string state:
\be\label{onshell}
q_i^2=-4/\ap\ \ \ ,\ \ \ i=1,2\ .
\ee

On the other hand the corresponding real open string planar and non--planar cylinder integrals are 
\begin{align}
A_{2;1}^{p}&:=\int_0^1 d\xi\  \fc{\theta_1\lf(\xi,\tau\ri)^2}{\eta^6}=-\fc{\theta_2(2\tau)}{\eta^6}\ ,\label{massopen}\\
A_{2;1}^{np}&:=\int_0^1 d\zeta\  \fc{\theta_4\lf(\zeta,\tau\ri)^2}{\eta^6}=\fc{\theta_3(2\tau)}{\eta^6},\label{massopeni}
\end{align}
The  integrals \req{massopen} and \req{massopeni} describe the one--loop mass renormalization in $SO(32)$ open superstring theory \cite{Yamamoto:1987dk}.
%% By using theta function doubling identities we have:
%% $$\begin{array}{lcl}
%% \lf|\theta_3\lf(\fc{\tau}{2}\ri)\ri|^2+\lf|\theta_4\lf(\fc{\tau}{2}\ri)\ri|^2
% &=&2 \th_3(\tau)^2\\
%% &=&2 \lf[\;\lf|\theta_2(2\tau)\ri|^2+\lf|\theta_3(2\tau)\ri|^2\;\ri].\end{array}$$
Thus, the complex torus integral \req{Marcus} can be cast into the following quadratic form:
\be\label{GREATKLT}
\widehat M_{2;1}^{closed}=2\tau_2^{1/2}\;|A_{2;1}^{p}|^2+2\tau_2^{1/2}\;| A_{2;1}^{np}|^2\ .
\ee
Note, that this is a particular simple DC structure relating a one--loop closed string integral to a sum over squares  of open string integrals. 

To compute the complex integral \req{Marcus} one starts
by expressing the square of  the theta--functions \req{theta1} as
\begin{align*}
e^{2G^{(1)}(z,\tau)}&=\fc{1}{4}e^{-4\pi\tau_2y^2}\sum_{p_i\in\{\pm1\}}\sum_{N_i,M_i\in\IZ}(-1)^{N_0+M_0}\\[2mm]
&\times e^{2\pi i z N_0} e^{-2\pi i \bar z M_0}q^{\fc{1}{4}(N_0^2+N_1^2)}\; \bar q^{\fc{1}{4}(M_0^2+M_1^2)},
\end{align*}
with the four integers:
\be
\begin{array}{lcl}
N_{0,1}&=&\fc{p_1}{2}(2n_1+1)\pm\fc{p_2}{2}(2n_2+1),\\[3mm]
%% N_1&=\fc{p_1}{2}(2n_1+1)-\fc{p_2}{2}(2n_2+1)\ ,\\[1mm]
M_{0,1}&=&\fc{p_3}{2}(2m_1+1)\pm\fc{p_4}{2}(2m_2+1).
%% M_1&=\fc{p_3}{2}(2m_1+1)-\fc{p_4}{2}(2m_2+1).
\end{array}
\ee
Then, the real $x$--integration gives the level--matching condition:
\be\label{levelmatch}
N_0=M_0.
\ee
The resulting integer sums over both even and odd $N_0$ can be used   to extend the real $y$--integration to a Gaussian integral leaving the integer sums with $N_1,M_1$ even or odd
subject to the solution \req{levelmatch} with $N_0$ even or odd, respectfully:
\be\label{MarcusRes}
\widehat M_{2;1}^{closed}=\fc{2\tau_2^{1/2}}{|\eta|^{12}}\Bigg\{\sum_{ N_1,M_1\;  even}+\sum_{N_1,M_1\; odd}\Bigg\}\;q^{\fc{1}{4} N_1^2} \bar q^{\fc{1}{4} M_1^2}.  
\ee
Eventually, the above expression leads to \req{Marcus}. 

Actually, the open string amplitudes \req{massopen} and \req{massopeni} conspire with one--loop open string monodromy relations \cite{Hohenegger:2017kqy,Tourkine:2016bak} as
\begin{align}
A_{2;1}^{p}&=-\tilde A_{2;1}^{p}\ ,\\
A_{2;1}^{np}&=\tilde A_{2;1}^{np}
\end{align}
giving rise to the  additional objects
\begin{align}
\tilde A_{2;1}^{p}&:=-\int_0^1 d\zeta\  \fc{\theta_4\lf(\zeta,\tau\ri)^2}{\eta^6}\ e^{2\pi i z}\;q^{-\tfrac{1}{4}},\label{Massopeni}\\
\tilde A_{2;1}^{np}&:=-\int_0^1 d\xi\  \fc{\theta_1\lf(\xi,\tau\ri)^2}{\eta^6}\ e^{2\pi i z}\;q^{-\tfrac{1}{4}}\ ,\label{Massopen}
\end{align}
with position dependent phases introduced in \cite{Hohenegger:2017kqy}.
As a consequence we may also write \req{GREATKLT} as
\be\label{newGREATKLT}
\widehat M_{2;1}^{closed}=2\tau_2^{1/2}\;|A_{2;1}^{p}|^2+2\tau_2^{1/2}\;|\tilde A_{2;1}^{np}|^2\ .
\ee

It is interesting to note, that the integrand
of \req{Marcus} has a $\IZ_2$--symmetry $z\ra-z$, i.e. it is sufficient to only integrate over 
a cylinder world--sheet $\cal C$.
Hence, it is instructive to split the torus integral \req{Marcus} into two contributions from  cylinder integrals as
\be
\widehat  M^{closed}_{2;1}=\widehat M^{p}_{2;1}+\widehat M^{np}_{2;1}\ ,
\ee
with the two cylinder integrals
\begin{align}
\widehat M^{p}_{2;1}&=\int_{\cal C}d^2z\; e^{2G^{(1)}(z,\tau)}=\h\;\widehat M^{closed}_{2;1}\ ,\\
\widehat M^{np}_{2;1}&=\int_{\cal C}d^2z\; e^{2G_T^{(1)}(z,\tau)}=\h\;\widehat M^{closed}_{2;1}\ ,
\end{align}
which can either be directly computed or by the methods developed in \cite{Stieberger:2021daa}.
Above, we have the twisted bosonic one--loop Green function
\be\label{BG}
G^{(1)}_T(z,\tau)=\ln\lf|\fc{\theta_4(z,\tau)}{\theta_1'(0,\tau)}\ri|^2-2\pi\fc{(\Im z)^2}{\Im \tau},
\ee
with the even Riemann theta--function:
\be\label{theta4}
\theta_4(z,\tau)=\sum_{n\in\IZ}(-1)^n q^{\h n^2}\cos(2\pi n z).
\ee

Hereinafter,  we shall use the alternative  expression of \req{Marcus} in terms of  a loop momentum $\ell$
which manifestly splits the integrand into a holomorphic and anti--holomorphic sector ($d=10$):
\begin{align}
\Mc^{closed}_{2;1}&=\delta^{(d)}(q_1+q_2)\int_{\Fc_1}{d^2\tau}\int_{-\infty}^{+\infty} d^d\ell\; e^{-\pi\ap \tau_2\ell^2}\nonumber\\
&\times
\int_{\cal T}\!d^2z\; e^{-i\pi \ap\ell q_1(z-\bar z)}\lf|\fc{\theta_1(z,\tau)}{\theta_1'(0,\tau)}\ri|^4.\label{MarcusLoop}
\end{align}
In fact, integrating first over the torus coordinate $z$ and performing the sum over $N_0$ constrains the loop momentum as:
\be\label{constraintl}
\ell'q_1=0\ \mbox{with:\ }\ell'=\ell+\tfrac{1}{2} q_1N_0\ .
\ee
Then, the remaining loop momentum integral decouples and can be performed by introducing spherical Lorentzian coordinates \cite{Nomizu} along the axis $q_1$:
\be\label{Birman}
\int_{-\infty}^{+\infty} d^d\ell'\; e^{-\pi\ap \tau_2\ell'^2}\delta^{(d)}(\ell'q_1)=||q_1||^{-1}\;(\ap\tau_2)^{\h(1-d)}.
\ee
Altogether, this yields  \req{MarcusRes} in a different way thereby constraining the loop momentum 
as \req{constraintl}.
This result underpins the holomorphic anti--holomorphic factorization of the result \req{GREATKLT}.
Furthermore, as it can be anticipated from \req{Birman} that the constraint \req{constraintl} entails the additional $\tau_2^{1/2}$--factors in \req{GREATKLT} and \req{GREATKLTEllmau}.

A similar discussion can be lead for the torus integral:
\be
\widehat {\frak M}^{closed}_{2;1}:=\int_{\cal T}d^2z\; e^{G^{(1)}(z,\tau)}=2\;\tau_2^{\h}\lf|\fc{1}{\eta^3}\ri|^{2}.\label{MarcusEllmau}
\ee
Similar to \req{massopen} and \req{massopeni}  we may introduce the following open string integrals:
\begin{align}
{\frak A}_{2;1}^{p}&:=\int_0^1 d\xi\  \fc{\theta_1\lf(\xi,\tau\ri)}{\eta^3}=\fc{2}{\pi}
\fc{q^{\tfrac{1}{8}}}{\eta^3}\sum_{n\in{\IZ}}(-1)^n\fc{q^{\h(n+1)n}}{2n+1}\ ,\label{massopenEllmau}\\
{\frak A}_{2;1}^{np}&:=\int_0^1 d\zeta\  \fc{\theta_4\lf(\zeta,\tau\ri)}{\eta^3}=\fc{1}{\eta^3}.\label{massopeniEllmau}
\end{align}
Hence, we have the following DC relation:
\be\label{GREATKLTEllmau}
\widehat {\frak M}_{2;1}^{closed}=2\;\tau_2^{1/2}\;|\frak A_{2;1}^{np}|^2=2\;\tau_2^{1/2}\;|\tilde{\frak A}_{2;1}^{np}|^2\ .
\ee
In addition, we have the objects
\begin{align}
\tilde {\frak A}_{2;1}^{p}&:=-i\int_0^1 d\zeta\  \fc{\theta_4\lf(\zeta,\tau\ri)}{\eta^3}\ e^{\pi i z}\;q^{-\tfrac{1}{8}}\label{MassopeniEllmau}\\
&=\fc{1}{\pi}
\fc{q^{-\tfrac{1}{8}}}{\eta^3}\sum_{n\in{\IZ}}(-1)^nq^{\h n^2}\lf(\fc{1}{2n+1}-\fc{1}{2n-1}\ri),\nonumber\\
\tilde {\frak A}_{2;1}^{np}&:=-i\int_0^1 d\xi\  \fc{\theta_1\lf(\xi,\tau\ri)}{\eta^3}\ e^{\pi i z}\;q^{-\tfrac{1}{8}}=\fc{1}{\eta^3}\ ,\label{MassopenEllmau}
\end{align}
which furnish the following open string monodromy relation \cite{Hohenegger:2017kqy}
\be\label{Scheppach}
{\frak A}_{2;1}^{p}-\tilde {\frak A}_{2;1}^{p}=2\;B_1\ ,
\ee
with the following boundary term \cite{Hohenegger:2017kqy}:
\be\label{Stihl}
B_1=\int_0^{\tau/2} dz\; \fc{\theta_1(z,\tau)}{\eta^3}\ .
\ee
Interestingly, as a side remark the relation \req{Scheppach} demonstrates the importance of the boundary term \req{Stihl} derived in \cite{Hohenegger:2017kqy}. This fact has also been stressed  in \cite{Casali:2019ihm}.

Finally, we shall mention, that expanding the exponential in the integrands of \req{Marcus} and \req{MarcusEllmau}
yields two--point modular graph functions \cite{Green:2008uj}
\be
D_k(\tau)=\int_{\cal T}d^2z\;G^{(1)}(z,\tau)^k,
\ee
e.g. $D_1\!=\!0,D_2\!=\!E(2,\tau)$, with the non--holomorphic Eisenstein series $E(s,\tau)$. Likewise, expanding the integrand of  \req{massopen} and \req{massopeni} yields two--vertex $B$-- and $A$--cycle holomorphic graph functions \cite{Broedel:2018izr}, respectively. Thus, our relations \req{GREATKLT} and \req{GREATKLTEllmau} are suited to generate relations between elliptic multiple zeta values and their single--valued objects, cf. also \cite{Zagier:2019eus}.

\begin{figure}[H]
%\centering
\hskip-0.5cm \includegraphics[width=0.5\textwidth]{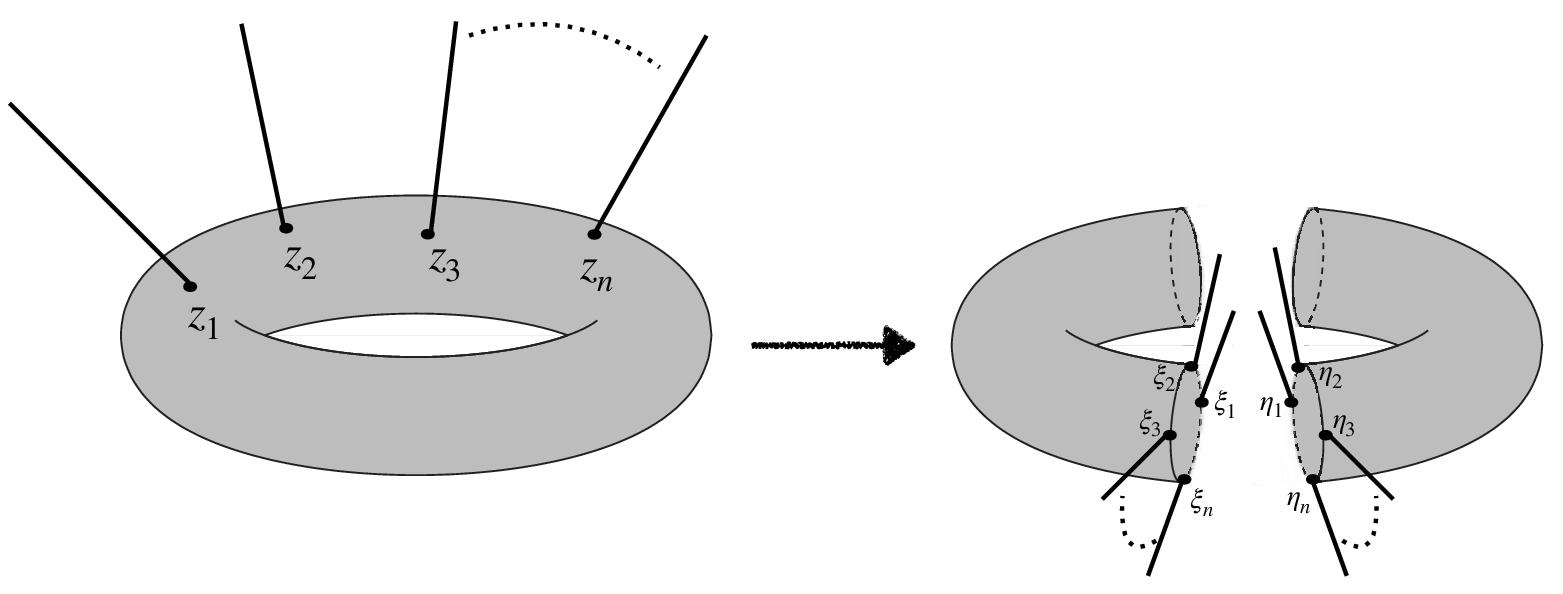}
\caption{Splitting the torus $n$--point amplitude into two cylinder amplitudes.}
\label{KLTsquare}
\end{figure}

Note, that \req{GREATKLT} and \req{GREATKLTEllmau} yield  KLT squaring identities at string one--loop in the spirit of \req{KLT4}. It would be very interesting to find more such examples of complex torus
integrals which can be written as squares of open string amplitudes in the spirit of \req{KLT4}. 
For generic $n$ one may expect   a splitting of the torus world--sheet into a double of cylinder world-sheets as depicted in Fig. \ref{KLTsquare}
\noindent
On the other hand, one--loop closed string amplitudes with logarithmic branch cuts in their low--energy expansion may not be simple squares of corresponding open string amplitudes.

Actually, a generalization of the single complex torus integrals \req{Marcus} and 
\req{MarcusEllmau} represents the  complex version of the Riemann--Wirtinger integral with non--integer powers of $\theta_1$ \cite{ghazouani2016moduli}. After proper implementing Riemann bilinear relations for complex conjugated (co)cycles its DC structure should be expressible  in terms of intersection numbers of twisted (co)homology classes at genus one~\cite{goto2023intersection}.

In the following we   discuss 
what DC structure to expect in the generic one--loop string case for multiple complex torus integrations.

\section{String one--loop double copy}
\def\Tc{{\cal T}}
\def\If{\frak I}

In string theory DC structures and numerators have been elaborated at tree--level for the massless case  in \cite{Mafra:2011kj,Broedel:2013tta}
 and for the massive case in \cite{Guillen:2021mwp,Lust:2023sfk}.
 The foundation  of these relations are the tree--level KLT relations \cite{Kawai:1985xq} and 
only recently a one--loop generalization thereof has been derived \cite{Stieberger:2022lss}. 
As in the tree--level case holomorphic properties of the string world--sheet are crucial to 
find such a relation. For this Cauchy's theorem is applied to study  monodromies and deformations of contours. The various steps 
are rather involved and will not be displayed here as they have been worked out in quite detail in  \cite{Stieberger:2022lss}. In contrast, here after  only briefly sketching the result we shall put emphasis on both its geometric impact  and working examples.

The one--loop string torus amplitude with $n$  closed oriented strings is given by
\be\label{fullclosed}
\Mc^{closed}_{n;1}(q_1,\ldots,q_n)\!=\!\h g_{c}^n\;\delta^{(d)}\!\lf(\sum_{r=1}^{n}q_r\ri)\!\!\int_{\Fc_1} \!\!\fc{d^2\tau}{\tau_2} \ M^{closed}_{n;1}\ ,
\ee
with the closed string coupling constant $g_c$ and the integrand
\be
M^{closed}_{n;1}=V_{CKG}^{-1}(\Tc)\!\left(\!\int_{\Tc}\prod_{s=1}^{n}d^2z_s\!\right)I(\{z_s,\bar z_s\})Q(\{z_s,\bar z_s\};\tau),\label{Start}
\ee
with some doubly--periodic  function $Q$ comprising possible kinematical factors. Generically, the latter assumes the form $Q\!=\!\tau_2^{1-d/2}Q_L(\tau)Q_R(\bar\tau)$. Furthermore, we have the integrand:
\begin{align}
I(\{z_s,&\bar z_s\})=\prod_{1\leq r<s\leq n}\lf[\fc{\theta_1(z_s-z_r,\tau)}{\theta_1'(0,\tau)}\ri]^{\h\ap q_sq_r}\label{ClosedString} \\
&\times\lf[\fc{\bar \theta_1(\bar z_s-\bar z_r,\bar \tau)}{\bar\theta_1'(0,\bar \tau)}\ri]^{\h\ap q_sq_r}\; \prod_{r,s=1\atop r<s}^{n} e^{-\fc{\pi\ap}{\tau_2}q_{r}q_{s}\Im(z_r-z_s)^2}.\nonumber
\end{align}
Note, that due to the lack of holomorphic double periodic functions on the torus we are dealing with quasi--periodic functions \req{BG} with non--harmonic contributions. As a consequence there is no holomorphic/anti--holomorphic factorization in contrast to the Virasoro--Shapiro amplitude \req{closed4}. Similar to \req{MarcusLoop} we introduce the loop momentum $\ell$ to holomorphically factorize the integrand  as~\cite{DHoker:1988pdl}:
\begin{widetext}
\bea
(\ap \tau_2)^{-d/2}\  I(\{z_s,\bar z_s\})&=&\ds\int_{-\infty}^\infty d^d\ell\; \exp\Big\{-\pi\ap \tau_2\ell^2-\pi i\ap\ell\sum_{r=1}^{n}
q_r(z_r-\bar z_r)\Big\}\\
&\times&\ds\prod_{1\leq r<s\leq n}\theta_1(z_s-z_r,\tau)^{\h\ap q_sq_r}\ 
\theta_1(\bar z_s-\bar z_r,\tau)^{\h\ap q_sq_r}.\label{Loopf}
 \eea
%\end{widetext}
To split each complex $z_t$--integration into a pair of real integrations   one now proceeds like in the tree--level case~\cite{Kawai:1985xq}
by considering contours in the complex plane at the cost of introducing phase factors.
After defining the parameterization 
$z_t=\sigma_t^1+i\sigma_t^2,\;t=1,\ldots,n$ with 
$\si_t^1\in(0,1)$ and $\si^2_t\in(-\frac{\tau_2}{2},\fc{\tau_2}{2})$
%% investigate monodromies in the complex torus $\sigma^2_t\in \Tc$ and consider closed loop
%% \be
%% \Gamma=C_1\cup C_2 \cup C_1'\cup C_2'
%% \ee
%% \begin{figure}[H]
%% \centering
%% \includegraphics[width=0.3\textwidth]{SigmaT.eps}
%% \caption{Closed contours in the complex $\sigma^2_t$--plane and branch points.}
%% \label{regge2}
%% \end{figure}
%% \noindent
for $\Re(\tau)=0$ we may consider some closed contour in the complex $\si^2_t$--plane and express the integration along the real axis $\si^2_t\in(-\frac{\tau_2}{2},\fc{\tau_2}{2})$ as some integral along the imaginary axis $\si_t^2\in(-i,0)$. This way  each complex $z_t$--integration   is traded  into a pair  of real integrations w.r.t. \cite{Stieberger:2022lss}
\be
\xi_t=\si^1_t+\tilde \si^2_t\ \ \ ,\ \ \
\eta_t=\si^1_t-\tilde \si^2_t,\label{newcoords}
 \ee
subject to some splitting function $\Psi$ to be specified below and some phases
%\begin{widetext}
\be\label{Phases}
\Pi_q(r,s):=\Pi(\xi_s,\xi_r,\eta_s,\eta_r; q_rq_s)=e^{\h\pi i\ap q_rq_s (1-\theta[(\xi_r-\xi_s)(\eta_r-\eta_s)])}
\ee
rendering the integrand of \req{Start} to be single--valued along $\xi_s,\eta_s\!\in\!(0,1)$. Eventually,  inserting the parameterization \req{newcoords} into the latter for $\Re(\tau)=0$  we obtain~\cite{Stieberger:2022lss}:
\bea
\ds M^{closed}_{n;1}(q_1,\ldots,q_n)&=&\ds\fc{1}{2}\; \delta^{(d)}\Bigg(\sum_{i=1}^n q_i\Bigg) \Bigg(\fc{i}{2}\Bigg)^{n-1} 
  \int_{-\infty}^\infty d^d\ell\ e^{-\pi\ap \tau_2\ell^2}\\[3mm]
&\times&\ds \int_0^1\Bigg(\prod_{r=1}^{n-1}d\xi_r\Bigg)  \int_0^1\Bigg(\prod_{r=1}^{n-1}d\eta_r\Bigg) \Bigg(\prod_{t=1}^{n-1}\Psi(\xi_t,\eta_t;\ell)\Bigg)\If_{n+2;0}(\ell) 
\Bigg(\prod_{r<s}^n\Pi_q(r,s)\Bigg) \tilde \If_{n+2;0}(\ell).\label{KLTone}
\eea
The objects in \req{KLTone} represent  specific integrands of (planar) one--loop open string amplitudes (with $g_c=g_o^2$): 
\begin{align}
\If_{n+2;0}(\ell)&=g_o^n\; \Bigg(\prod_{r,s=1\atop r<s}^{n} 
|\theta_1(\xi_s-\xi_r,\tau)|^{\h\ap q_sq_r}\Bigg)\;e^{-i\pi\ap \ell\sum\limits_{r=1}^{n}q_r\xi_r}\;Q_L(\tau,\{\xi_s\}),\\[-2mm]
\tilde \If_{n+2;0}(\ell)&=g_o^n\;  \Bigg(\prod_{r,s=1\atop r<s}^{n} 
 |\theta_1(\eta_s-\eta_r, \tau)|^{\h\ap q_sq_r}\Bigg)\; e^{i\pi\ap \ell\sum\limits_{r=1}^{n}q_r\eta_r}\; Q_R(\ov\tau,\{\eta_s\}).
 \end{align}
%% with $n\!+\!2$ open string insertions and three marked points and  the open string momenta
%% \be
%% p_i=\h q_i,\ i=1,\ldots,n\ \ \ ,\ \ \ p_0=\h\;\ell\ \ \ ,\ \ \ p_{n+1}=-\h\;\ell\ .
%% \ee
\end{widetext}

\begin{figure}[H]
\centering
\includegraphics[width=0.35\textwidth]{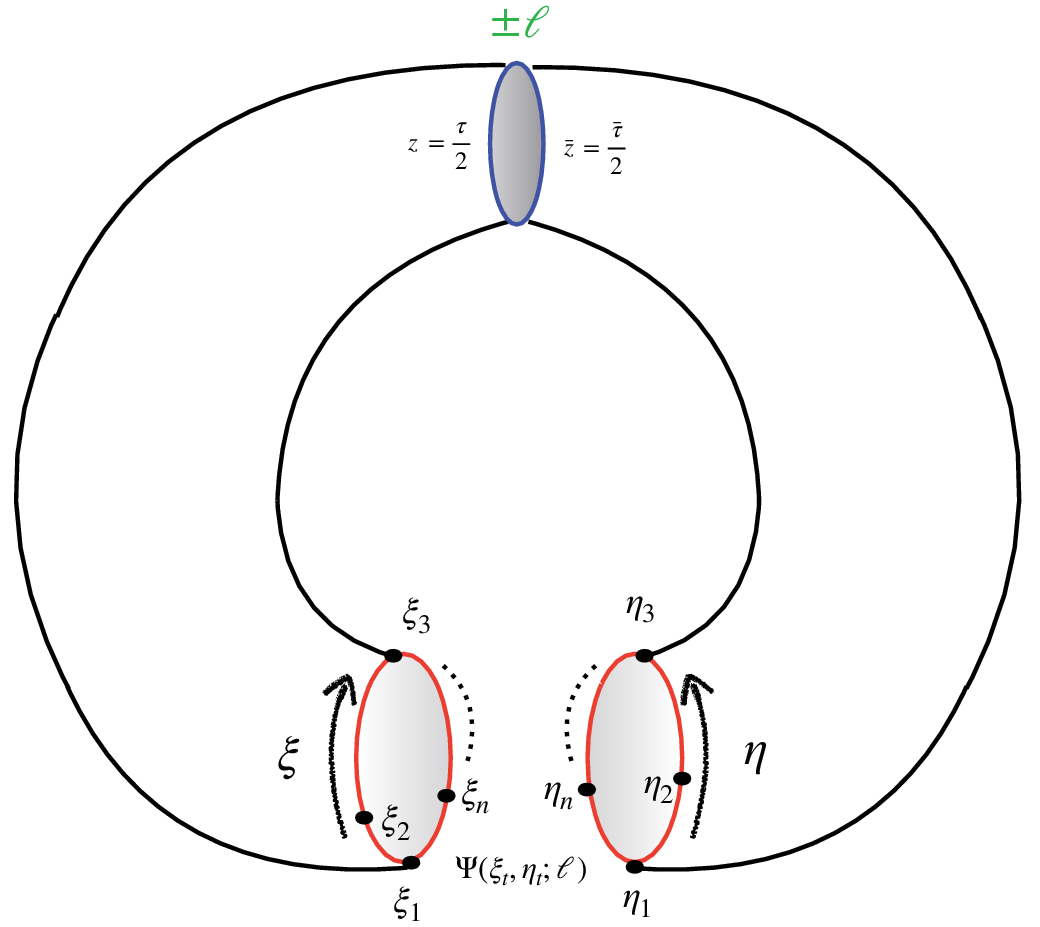}
\caption{Slicing the torus along the $A$--cycle into one cylinder with a closed string insertion of  momentum  $\pm\ell$.}
\label{NonPlanar}
\end{figure}
\noindent 
As proposed above, the final result \req{KLTone} furnishes a splitting of complex torus integrations into holomorphic and anti--holomorphic sectors just like at tree--level. However, the main difference at one--loop is the splitting function $\Psi$, which accounts for the change of torus coordinates \req{newcoords}. The underlying world--sheet of the expression \req{KLTone} can be interpreted as a non--planar cylinder   with a closed string insertion, cf. Fig.~\ref{NonPlanar}. More precisely, the one--loop torus is sliced along the $A$--cycle with $2n$ open string positions 
$\xi_i$ and $\eta_j$ located along the two boundaries, respectively resulting in a non--planar one--loop cylinder 
configuration.
The details of the cutting procedure is governed by the following splitting  function
\be\label{EllisJetzer}
\Psi(\xi_t,\eta_t;\ell)=\fc{(1+e^{-\pi i\ap\ell q_t})}{(1-e^{-2\pi i \ell q_t})} \;
 e^{-\pi i\ap\ell q_t\;\theta(\eta_t-\xi_t)}
 \ee
originating  from the change of coordinates \req{newcoords} along the two boundaries.
The function \req{EllisJetzer} intertwines the real integrations $\xi_s,\eta_s$ with the phase factor $\Pi_q$. 
Furthermore, the splitting function $\Psi$ essentially subjects level matching conditions 
to the left-- and  right--movers, which will be evidenced  below.

\begin{figure}[H]
\centering
\includegraphics[width=0.4\textwidth]{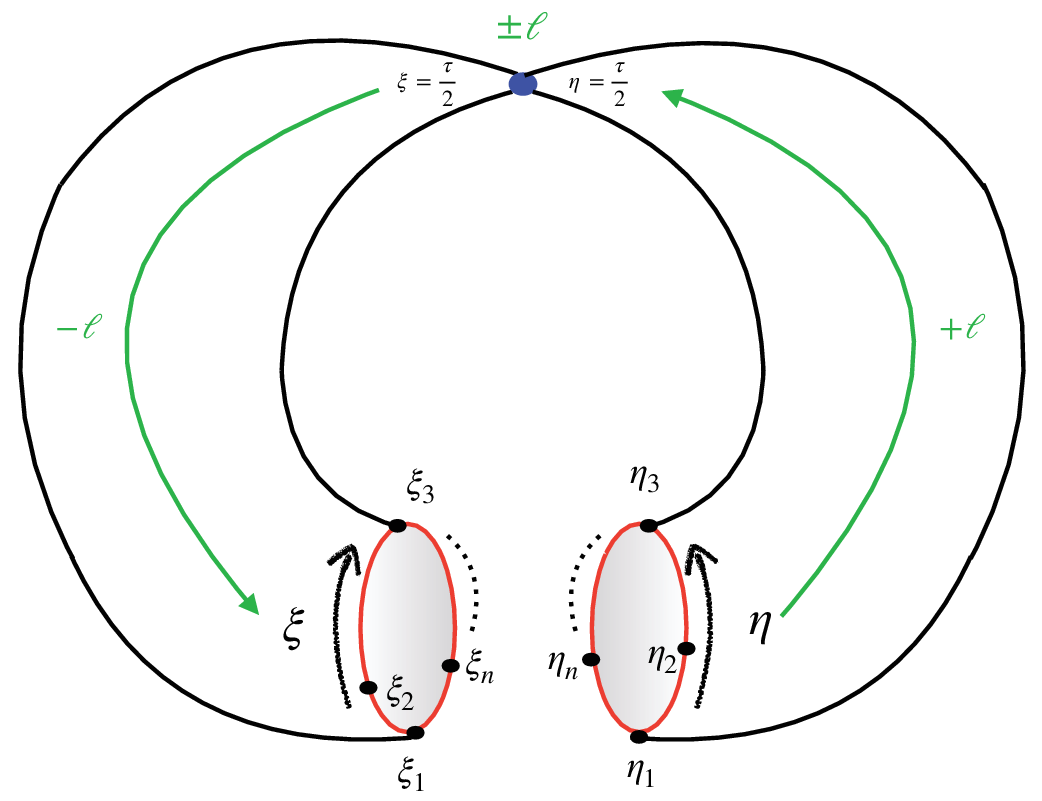}
\caption{Slicing the torus into two cylinders connected by a closed string node exchanging the loop momentum $\pm\ell$.}
\label{Zipfel}
\end{figure}

In the large complex structure limit $\tau\ra i\infty$ the closed string becomes a node connecting
two degenerating cylinders. 
In this limit the torus is pinched to a node along the $B$--cycle and  the diagram Fig.~\ref{NonPlanar} turns into a product of two disk diagrams each with a single closed string insertion at the node, cf. Fig.~\ref{Zipfel}.
\noindent 
\noindent
This limit has thoroughly  been worked out in \cite{Stieberger:2022lss}. In particular, the field--theory limit $\ap\!\ra\!0$ is governed by the large complex structure limit $\tau_2\!\ra\!\infty$ of the integrand of~\req{fullclosed} and exhibits a similar structure than the field--theory DC formula \cite{Stieberger:2022lss}
\begin{align}
{\cal M}^{grav}_{n;1}&\simeq \h\delta^{(d)}\Big(\sum\limits_{i=1}^n q_i\Big) \int \fc{d^d\ell}{\ell^2}\label{Einserkogel}\\
&\sum_{\si,\rho\in S_{n-1}}\!\! A_{n+2;0}(+\ell,\sigma(1,\ldots,n-1),n,-\ell)\nonumber\\
&\times S[\sigma|\rho]_\ell\;\tilde A_{n+2;0}(+\ell,\rho(1,\ldots,n-1),-\ell,n),\nonumber
\end{align}
involving the loop momentum $\ell$ and the  (off--shell) $n+2$--point tree--level gluon amplitudes 
in the forward limit
\be\label{gluonn+2}
A_{n;1}(1,\ldots,n)\!=\!\int\fc{d^d\ell}{\ell^2}\!\!\!\! \sum_{\gamma\in cyc(1,\ldots,n)}\!\!\!\!\!
 A_{n+2;0}(+\ell,\gamma(1,\ldots,n),-\ell),
\ee
with the external momenta $\pm\ell$ \cite{Geyer:2015bja,Geyer:2015jch}.
Furthermore, there is the field--theory kernel \req{kernel}, with $S[\sigma|\rho]_\ell:=\lim_{\ap\ra0}(\pi\ap)^{1-n}\Sc[\sigma|\rho]_\ell$. In this formulation the loop momentum $\ell$ is identified with a light--like external momentum of a tree--level amplitude $A_{n+2;0}$. The expression \req{Einserkogel} has formerly   been conjectured  in \cite{He:2016mzd}.

Let us now return to the example \req{Marcus} and its loop momentum description \req{MarcusLoop}  \cite{cyc}. 
For this case in the general expression \req{Start} we have $z_1\!=\!z,z_2\!=\!0$ and \req{onshell}.
Then \req{KLTone} becomes: 
\begin{widetext}
\bea\label{Wasnun}
\ds M_{2;1}^{closed}(q_1,q_2)&=&\ds\fc{1}{4}\fc{i}{2}\!\!\!\sum_{p_i\in\{\pm1\}}\int_{-\infty}^\infty d^d\ell\; e^{-\pi\ap\tau_2\ell^2}\;\sum_{N_i,M_i\in\IZ}(-1)^{N_0+M_0}q^{\fc{1}{4}(N_0^2+N_1^2)}\; 
\bar q^{\fc{1}{4}(M_0^2+M_1^2)}\\[5mm]
&\times&\ds\int_0^1\!\!d\xi\int_0^1\!\!d\eta\; \Psi(\xi,\eta;\ell)\;
 e^{2\pi i \xi (N_0-\fc{\ap}{2}\ell q_1)} e^{-2\pi i \eta (M_0-\fc{\ap}{2}\ell q_1)}.
\eea
After performing the real $\xi,\eta$--integrations we evidence  the imposition of the level--matching condition \req{levelmatch}:
\be\int_0^1\!\!d\xi\int_0^1\!\!d\eta\; \Psi(\xi,\eta;\ell)\;
 e^{2\pi i \xi (N_0-\fc{\ap}{2}\ell q_1)} e^{-2\pi i \eta (M_0-\fc{\ap}{2}\ell q_1)}=-\fc{\delta(M_0-N_0)}{2\pi i(M_0-\fc{\ap}{2}\ell q_1)}.
\ee
%\begin{widetext}
After shifting the loop momentum by $\ell=\ell'-\h q_1M_0$ in accord with \req{constraintl} we may 
cast \req{Wasnun} into the following form:
\be\label{Stock1} 
M_{2;1}^{closed}=2\int_{-\infty}^\infty d^d\ell'\;
\fc{e^{-\pi\ap\tau_2\ell'^2}}{2\pi\ap\ell'q_1}\;\lf(e^{\pi\ap \ell'q_1\tau_2}-e^{-\pi\ap \ell'q_1\tau_2}\ri)\!\lf\{\lf|\fc{\theta_2(2\tau)}{\eta^6}\ri|^2\!\!\!\sum_{N_0\; \mbox{\small even}}e^{\pi\ap \ell'q_1\tau_2N_0}
+\lf|\fc{\theta_3(2\tau)}{\eta^6}\ri|^2\!\!\!\sum_{N_0\; \mbox{\small odd}}e^{\pi\ap \ell'q_1\tau_2N_0}\ri\}\!. 
\ee
On the other hand, the corresponding expression from \req{MarcusLoop} 
yields:
\be\label{Stock2}
 M_{2;1}^{closed}=2\int_{-\infty}^\infty d^d\ell'\;
\fc{e^{-\pi\ap\tau_2\ell'^2}}{2\pi\ap\ell'q_1}\;\lf(e^{2\pi\ap \ell'q_1\tau_2}-1\ri)\lf\{\lf|\fc{\theta_2(2\tau)}{\eta^6}\ri|^2\!\!\!\sum_{N_0\; \mbox{\small even}}e^{\pi\ap \ell'q_1\tau_2N_0}
+\lf|\fc{\theta_3(2\tau)}{\eta^6}\ri|^2\!\!\!\sum_{N_0\; \mbox{\small odd}}e^{\pi\ap \ell'q_1\tau_2N_0}\ri\}. 
\ee
\end{widetext}
\ \\

\noindent
The last two expressions \req{Stock1} and \req{Stock2} involve infinite sums
\begin{align}
\sum_{N_0\; \mbox{\small even}}e^{\pi\ap \ell'q_1\tau_2N_0}&=\tau_2^{-1}\;\delta(\ap\ell' q_1)\;,\\
\sum_{N_0\; \mbox{\small odd}}e^{\pi\ap \ell'q_1\tau_2N_0}&=\tau_2^{-1}\;\delta(\ap\ell' q_1)\;,
\end{align}
and agree subject to the delta--function support \req{constraintl} leading to
\begin{align}
M_{2;1}^{closed}&=2\int_{-\infty}^\infty d^d\ell'\;
e^{-\pi\ap\tau_2\ell'^2}\ \delta(l'q_1)\nonumber\\
&\times\lf\{\lf|\fc{\theta_2(2\tau)}{\eta^6}\ri|^2+\lf|\fc{\theta_3(2\tau)}{\eta^6}\ri|^2\ri\}. 
\end{align}
This is the result stemming from the direct computation \req{MarcusLoop} and in agreement with \req{Marcus}. A similar check can be 
done for the example \req{MarcusEllmau}.

\section{Concluding remarks}

 In  Eqs. (II.17) and (II.37) we have presented examples of  fully--fledged string--one loop double copies for the first time in the literature. In addition, their underlying one--loop string monodromies are discussed.

For $\Re\tau=0$ our result \req{KLTone} generalizes the tree--level KLT  relations to one--loop and it can be applied for  both the massless and massive case -- with or without supersymmetry. The result \req{KLTone} is the first generalization of the tree--level KLT relations to
loop--level, which has a great potential impact on the double copy
relations and all their uses.
 As a consequence in the field--theory limit our relations capitalize  solid one--loop gauge--gravity relations including loop--level color kinematics duality.  Generalization of  \req{KLTone} to $\Re\tau\neq 0$ is very interesting. This task requires extending the analytic continuation of complex vertex operator positions to non--rectangular tori.
 
Complementary, in some  recent work  the imaginary part of a one--loop string amplitude is computed by considering unitary cuts of the string world--sheet and including massive states \cite{Eberhardt:2022zay}.
 At tree--level there are further relations between closed and open string world--sheet diagrams due to the single--valued projection, cf. for \cite{Stieberger:2016xhs} a review. Furthermore, a kind of opposite question is when starting from a single--valued amplitude  and asking how the latter can be related  to a pair of amplitude expressions with multi--valued coefficients, cf. interesting work~\cite{Baune:2023uut}.

\noindent
{\bf Acknowledgments:}
I wish to thank  Johannes Broedel and Pouria Mazloumi for interesting discussions.

%\vskip-0.75cm
\nocite{*}
\bibliography{KLTone}% Produces the bibliography via BibTeX.

\begin{thebibliography}{10}
\bibitem{cyc}
We have the relation $\left.M^{closed}_{n;1}=\tau_2^{-d/2}\widehat
  M^{closed}_{n;1}\right|_{\Re(\tau)=0}$.
\end{thebibliography}
\bibliographystyle{h-physrev5}

\end{document}